\documentclass[conference,10pt]{IEEEtran}
\usepackage{cite}
\usepackage{comment}
\usepackage{amsmath,amssymb,amsfonts,bm}
\usepackage{graphicx}
\usepackage[nolist]{acronym}
\usepackage{textcomp}
\usepackage[table,dvipsnames]{xcolor}
\usepackage{tikz}
\usetikzlibrary{positioning,shapes,matrix,fit,backgrounds,arrows,chains,patterns}
\usepackage{pgfplots} 
\usepgfplotslibrary{fillbetween} 
\usepackage[outline]{contour}
\usepackage[commentColor=black]{algpseudocodex}
\tikzset{algpxIndentLine/.style={draw=black}}
\algrenewcommand{\alglinenumber}[1]{\bfseries\footnotesize #1}
\algrenewcommand{\textproc}{}
\algrenewcommand{\algorithmicrequire}{\textbf{Input:}}
\algrenewcommand{\algorithmicensure}{\textbf{Output:}}
\usepackage{algorithm}
\usepackage{caption} 
\usepackage{subcaption} 
\usepackage{hyperref}

\def\BibTeX{{\rm B\kern-.05em{\sc i\kern-.025em b}\kern-.08em
    T\kern-.1667em\lower.7ex\hbox{E}\kern-.125emX}}

\newcommand\copyrighttext{%
    \footnotesize \textcopyright 2024 IEEE. Personal use of this material is permitted. Permission from IEEE must be obtained for all other uses, in any current or future media, including reprinting/republishing this material for advertising or promotional purposes, creating new collective works, for resale or redistribution to servers or lists, or reuse of any copyrighted component of this work in other works. DOI: \href{https://ieeexplore.ieee.org/document/10693977}{10.1109/SPAWC60668.2024.10693977}
}
\newcommand\copyrightnotice{%
\begin{tikzpicture}[remember picture,overlay]
\node[anchor=south,yshift=10pt] at (current page.south) {\fbox{\parbox{\dimexpr\textwidth-\fboxsep-\fboxrule\relax}{\copyrighttext}}};
\end{tikzpicture}%
}

\begin{document}

\begin{acronym}
  \acro{ML}{Machine Learning}
  \acro{AI}{Artificial Intelligence}
  \acro{NN}{Neural Network}
  \acro{DNN}{Deep Neural Network}
  \acro{BP}{Belief Propagation}
  \acro{NBP}{Neural Belief Propagation}
  \acro{GNBP}{Gated Neural Belief Propagation}
  \acro{ECC}{Error Correcting Codes}
  \acro{RL}{Reinforcement Learning}
  \acro{LDPC}{Low Density Parity Check}
  \acro{DL}{Deep Learning}
  \acro{BLER}{Block Error Rate}
  \acro{SGD}{Stochastic Gradient Descent}
  \acro{SNR}{Signal to Noise Ratio}
  \acro{AWGN}{Additive White Gaussian Noise}
  \acro{LLR}{Log Likelihood Ratio} 
  \acro{RNN}{Recurrent Neural Network}
  \acro{BCE}{Binary Cross-Entropy}
  \acro{SGD}{Stochastic Gradient Descent}
\end{acronym}

\title{Semantic Communications Services within Generalist Operated Networks}

\author{
    \IEEEauthorblockN{Quentin Lampin, Louis-Adrien Dufrène, Guillaume Larue}
    \IEEEauthorblockA{Orange Research, Meylan, France\\
    Email: \{quentin.lampin, louisadrien.dufrene, guillaume.larue\}@orange.com}
}

\maketitle

\copyrightnotice
\begin{abstract}
This paper addresses the challenge of integrating semantic communication principles into operated networks, traditionally optimized based on network-centric metrics rather than application-specific needs. Operated networks strongly adhere to the principle of ``separation of concerns", which emphasizes a clear distinction between network operation and application. Despite the initial perceived incompatibility between semantic communication and the principles of operated networks, this paper provides solutions to reconcile them. The foundations of these solutions include the adoption of non-arbitrary semantic representations as a standard encoding for communications, the establishment of a standard interface between the application and network, and a dedicated network control plane. These enable the application to describe the data typology and the nature of the task, and to agree upon a transmission scheme tailored to the supported task. Through three scenarios involving an application transmitting text representations, we illustrate the implementation of the proposal and demonstrate the potential of the approach.
    
\end{abstract}


\section{Introduction}

    Early telecommunication networks emerged as highly specialized systems, tailored within the technological means of the time to an inevitably limited set of use cases. For instance, the Global System for Mobile Communications (GSM) network was primarily developed for efficient voice transmission, with the entire system optimized to fulfill this sole purpose. 
    
    Technological advancements and increasing demands for a variety of communication services, \emph{e.g.} text messaging, multimedia sharing, and internet access, prompted the evolution of networks into more adaptable structures. This led to the advent of generalist networks, relying on a universal data representation---bits---and based on the fundamental concept of ``separation of concerns": the applicative service, which understands the content and purpose of the data; and the network function, which is solely responsible for agnostic data delivery. 
    Modern generalist telecommunications networks, such as 4G and 5G, have been developed under this paradigm.

    A direct consequence of the ``separation of concerns" principle is that communication networks have been built around purely network's performance metrics such as latency, throughput, error rates, capacity, and energy consumption, rather than application's performance metrics. To ensure an application functions properly, it is necessary to translate the application's expectations into these network's metrics. As an illustration, a streaming application may stipulate requirements in form of a minimum bit rate and latency to ensure quality of service. Within these generalist networks, meeting the performance demands of specific service classes often comes at the expense of dedicated adaptations of network functions and equipment.    
    For instance, the categorization of 5G networks use cases into enhanced Mobile Broadband (eMBB), massive Machine Type Communications (mMTC), and Ultra-Reliable Low-Latency Communications (URLLC) reflects the need to tailor networks to particular use case families \cite{redcap}. Although a network is ultimately intended to serve specific application purposes, its design remains anchored to network-centric metrics, rather than those focused on the application. In real-world settings, this enduring ``separation of concerns" usually results in sub-optimal solutions with regard to application performance.

    In this context, semantic communications place the application objectives at center of the communication process, by systematically considering metrics related to the application's needs, rather than those focused on the network's performance. In the state of the art, this is often achieved by tying the application to the network, forming a unified system that is optimized in an end-to-end manner. The semantic representations to be transmitted are then tailored to fulfill the receiver's application goal \cite{beyondBits}. 
    However, the oversight regarding the enduring separation between application and network operations raises questions. Although this divide can be overcome in a private network or among services within the network---because they are usually managed by the same organization---this persistent separation presents a major obstacle to incorporating semantic communication methods into the most commonly used communication systems, namely operated networks.

\section{A Network Operator Perspective}

    \begin{figure*}[htbp]
        \centering  
        \begin{tikzpicture}[xscale=0.65,yscale=0.57,every node/.style={transform shape,font=\fontsize{12pt}{12pt}\selectfont}] 
            \contourlength{1.4pt}
            \tikzset{>=latex} 
            \linespread{1}
            %

            \node (ORIGIN) at (0,0) {};

            \path (ORIGIN.east) ++ (+0,0) coordinate (APP_TX_1);
            \node [draw,line width=.8pt,rounded corners,minimum height=1.6cm,align=center] (APP_TX) at (APP_TX_1) {Application\\ source};

            \path (APP_TX.south) ++ (0,-3) coordinate (OPERATOR_TX_1);
            \node [draw,line width=.8pt,rounded corners,minimum height=1.6cm,align=center] (OPERATOR_TX) at (OPERATOR_TX_1) {TX Endpoint};

            \path (OPERATOR_TX.east) ++ (+7,0) coordinate (CHANNEL_1);
            \node [draw,cylinder,line width=.8pt,minimum height=5cm,minimum width=1.5cm,align=center] (CHANNEL) at (CHANNEL_1) {Physical Network};

            \path (CHANNEL.east) ++ (+7,0) coordinate (OPERATOR_RX_1);
            \node [draw,line width=.8pt,rounded corners,minimum height=1.6cm,align=center] (OPERATOR_RX) at (OPERATOR_RX_1) {RX Endpoint};

            \path (OPERATOR_RX.north) ++ (+0,+3) coordinate (APP_RX_1);
            \node [draw,line width=.8pt,rounded corners,minimum height=1.6cm,align=center] (APP_RX) at (APP_RX_1) {Application\\ recipient};

            \path (APP_TX.east) ++ (+7,0) coordinate (S_CHANNEL_1);
            \node [xscale=0.95,yscale=0.85,cloud, draw,cloud puffs=15,cloud puff arc=120, aspect=3, inner ysep=0.0em] (S_CHANNEL) at (S_CHANNEL_1){Abstract Semantic Channel};

            \path (OPERATOR_TX.west) ++ (-0.5,+1.25) coordinate (OPERATOR_BOX_1);
            \path (OPERATOR_RX.east) ++ (+0.5,-1.5) coordinate (OPERATOR_BOX_2);    
            \path (CHANNEL.south) ++ (+0,-0.5) coordinate (OPERATOR_BOX_LABEL_1);
            \node [align=center] (OPERATOR_BOX_LABEL) at (OPERATOR_BOX_LABEL_1) {\textbf{OPERATOR SEMANTIC NETWORK PLATEFORM}};
            \node (OPERATOR_BOX) [draw,line width=0.5pt,dashed,rounded corners=15pt,fit = (OPERATOR_BOX_1) (OPERATOR_BOX_2)] {};

            \path (APP_TX.west) ++ (-0.5,+1.75) coordinate (APP_BOX_1);
            \path (APP_RX.east) ++ (+0.5,-1.25) coordinate (APP_BOX_2);    
            \path (S_CHANNEL.north) ++ (+0,+0.25) coordinate (APP_BOX_LABEL_1);
            \node [align=center] (APP_BOX_LABEL) at (APP_BOX_LABEL_1) {\textbf{APPLICATION}};
            \node (APP_BOX) [draw,line width=0.5pt,dashed,rounded corners=15pt,fit = (APP_BOX_1) (APP_BOX_2)] {};

            \draw[very thick,-latex,rounded corners] ([xshift=-0.25cm] APP_TX.south) -- ([xshift=-0.25cm] OPERATOR_TX.north) node [left,pos=0.5] {Semantic Payload};;
            \draw[very thick,latex-latex,dashed,rounded corners] ([xshift=+0.25cm] OPERATOR_TX.north) -- ([xshift=+0.25cm] APP_TX.south) node [right,pos=0.5] {Semantic Interface};

            \draw[very thick,-latex,rounded corners] ([yshift=+0.25cm] OPERATOR_TX.east) -- ([yshift=+0.25cm] CHANNEL.west) node [above,pos=0.5, align=center] {Data Plane\\Encoded Embeddings};
            \draw[very thick,latex-latex,dashed,rounded corners] ([yshift=-0.25cm] OPERATOR_TX.east) -- ([yshift=-0.25cm] CHANNEL.west) node [below,pos=0.5] {Control Plane};
            \draw[very thick,-latex,rounded corners] ([xshift=-0.25cm,yshift=+0.25cm] CHANNEL.east) -- ([yshift=+0.25cm] OPERATOR_RX.west) node [above,pos=0.5, align=center] {Data Plane\\Encoded Embeddings};            
            \draw[very thick,latex-latex,dashed,rounded corners] ([xshift=-0.25cm,yshift=-0.25cm] CHANNEL.east) -- ([yshift=-0.25cm] OPERATOR_RX.west) node [below,pos=0.5] {Control Plane};

            \draw[very thick,latex-latex,dashed,rounded corners] ([xshift=-0.25cm] OPERATOR_RX.north) -- ([xshift=-0.25cm] APP_RX.south) node [left,pos=0.5] {Semantic Interface};
            \draw[very thick,-latex,rounded corners] ([xshift=+0.25cm] OPERATOR_RX.north) -- ([xshift=+0.25cm] APP_RX.south) node [right,pos=0.5] {Semantic Payload};

            \draw[very thick,-latex,dotted,rounded corners] (APP_TX) -- (S_CHANNEL) node [above,pos=0.5,align=center,text width=2.5cm] {Transmitted Embeddings};
            \draw[very thick,-latex,dotted,rounded corners] (S_CHANNEL) -- (APP_RX) node [above,pos=0.5,align=center,text width=2.5cm] {Received Embeddings};




        \end{tikzpicture}
        \caption{Operated Semantic Network Architecture}
        \label{fig:OperatedSemanticNetwork}

    \end{figure*}
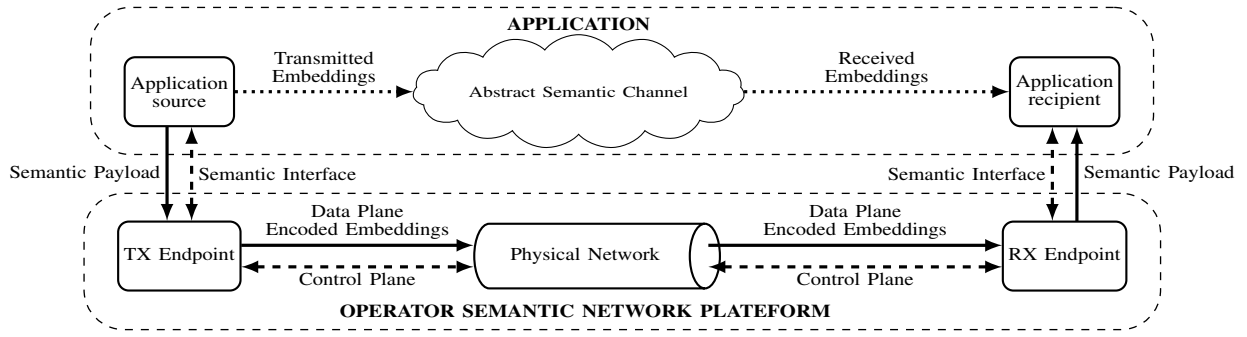
    
    Bridging the gap between application services and network operations is a key challenge in bringing semantic communication principles into operated networks. Applications are generally reluctant to expose private data and sensitive information to the network for optimization purposes; and likewise, network operators are cautious about exposing operational details. Coupled with interoperability requirements, this led to a preference for generic transmission mechanisms and the negotiation of Service Level Agreements (SLAs) that guarantee network performance based on network-centric Key Performance Indicators (KPIs), such as latency, error rate, and capacity, which do not require disclosure of business or operational data.
    
    Despite the difficulties posed by these limitations, we are confident that incorporating semantic communications principles into operated networks is possible by establishing a standard but adaptable framework. This framework, presented in Fig. \ref{fig:OperatedSemanticNetwork}, is articulated around three enablers: (i) non-arbitrary semantic representations, (ii) dedicated interfaces to negotiate semantic-centric SLAs and exchange associated information and metrics, and (iii) a semantic control plane to configure network functions accordingly.

    \subsection{Non-Arbitrary Semantic Representations}
    
    Moving towards semantic communication networks involves adopting a new, universal data representation that can be managed by Machine Learning (ML) models to meet diverse application-specific requirements. Embeddings, central in the recent observed advance in AI, are multi-dimensional vectors well-suited for this role. Multi-modal models like Google DeepMind's Gemini\cite{geminiteam2024gemini} have shown how embeddings can effectively manage different data types within a unified representation framework. Their ability to capture detailed data enables the creation of precise, application-specific encodings, leading to better optimization. Furthermore, embeddings are inherently compatible with the increasing number of AI-based applications, facilitating native communication between them. By relying on machine learning techniques, embeddings enable the automatic construction of information representations tailored for specific (communication) tasks, sidestepping the sub-optimality of arbitrary encodings w.r.t Shannon's information theory.

    Therefore, embeddings are poised to become a fundamental element of information in semantic networks, which are expected to specialize in transmitting this versatile data format, just as traditional networks specialized in transmitting bits.
            
    \subsection{Enhanced Application-Network Interface}
    
        To enable the network to offer semantic communication services, new standardized interfaces between the network and applications must be established, as depicted in Fig. \ref{fig:OperatedSemanticNetwork}. The degree to which source and destination applications share extra information with the network correlates with the network's capability to tailor its functions to the applications' specific needs. We outline five levels of progressive integration and complexity.

        \textbf{Level 0}: The network is unaware of the semantic significance of the data it handles. It is unable to provide a service beyond traditional SLAs and is solely responsible for agnostic binary data delivery to the recipient host. Although the application may handle tasks like compression on its semantic representations, it cannot tailor its encoding to network specifics such as the symbol mapping. This lack of cooperation subjects both the network and application to inefficiencies, \emph{e.g.} overuse of transmission resource and overall degradation of the service delivery.

        \textbf{Level 1}: The network becomes aware that embeddings are transmitted but is oblivious to their intrinsic properties or their application-level utilization. It enables the network and applications to negotiate the transport of data through generalist semantic reconstruction metrics that cater to a range of application tasks, like preserving the cosine similarity of embeddings during transmission. It helps the network to better mitigate any caused semantic distortions through generic optimizations, while reducing resource usage. From this level onwards, the network truly becomes an active participant in the semantic communication system, evolving into a ``semantic-aware network".
        
        \textbf{Level 2}: The application provides metrics and information on the embeddings, \emph{e.g.} embedding model or statistics on values, without task's specifics. The network exploits embeddings' intrinsic characteristics to better evaluate and minimize any caused distortion. Owing to the knowledge of the embedding model, SLAs specific to the embeddings model can be negotiated on reference datasets w.r.t reconstruction metrics.

        \textbf{Level 3}: The application supplies general information on its task. This may include the type of task performed, \emph{e.g.} classification, or nature of the encoded data, \emph{e.g.} customer text inquiries. This level of integration allows the network to exploit existing reference benchmarks and datasets relevant to the application use case which may serve as basis for the SLAs negotiation.
        
        \textbf{Level 4}: The application directly provide the network with application-specific metrics, tasks and embeddings. This level empowers the network to fine-tune its transmission strategies in an end-to-end manner, thereby forging a strong connection with semantic and purpose-driven communication. Shared information might include labeled embeddings, source data, application's accuracy feedback, etc.
        
        The aspects detailed above emphasize the imperative for a bespoke Semantic Interface that facilitates the exchange of information and metrics between the application and the network. This interface would also underpin the formulation of a novel category of SLAs, termed Semantic SLAs (SSLAs), which are predicated on metrics related to embeddings and/or specific to applications, diverging from conventional network-centric KPIs. The formalization of such SSLAs is certainly a challenge on its own, both technically due to the diversity of application tasks and associated metrics, and legally as the network is not able to guarantee performance levels that are affected by the application's internal workings.

    \subsection{Semantic Control Plane}
        
        The diverse data representations and associated network's processing, which are essential for meeting the application's requirements, necessitate negotiation and signaling between the involved network's nodes to ensure effective data transmission. Hence, a dedicated control plane is needed to manage this signaling and the configuration of models, KPIs, embedding processing... Since semantic communications usually rely on the use of ML to adapt and process the semantic representations, the control plane should also facilitate the transfer of training data, neural network models or contextual information, and more generally support distributed learning processes \cite{strinati2024goaloriented}.

\section{The case for semantic network integration}
\label{sec:system_model}

    {\input{figures/quantification_indexation}}

    In this section, we present the rationale for semantic-aware processing within operated networks, using the example of a network supporting an application that transmits text representations for processing, analysis, and storage. Specifically, we simulate a system where a conversational agent receives text messages and classifies customer inquiries. The dataset considered is the Bitext dataset \cite{bitex}.
    
    To showcase the potential of semantic-aware networks, we consider three scenarios:
    
    \textbf{Scenario 1}: Customer inquiries are transmitted as text in a conventional arbitrary text format.
            
    \textbf{Scenario 2}: Customer inquiries are encoded as sentence embeddings by the application. Those are transmitted in a standard arbitrary floating-point format. The network is unaware of the semantic nature of the representations being transmitted.
    
    \textbf{Scenario 3}: Customer inquiries, encoded as sentence embeddings by the application, are submitted to the network for processing and transmission in the form of binary datagrams. The sentence embeddings are processed by the network with respect to a reference benchmark task and dataset.
    
    For this illustration, the transmission is carried over a logical channel between two hosts modeled as a Binary Symmetric Channel (BSC) with symbol error probability $p_e$.
    
    \subsection{Arbitrary data format over a non-semantic network} 
        In scenario 1 and 2, the network supports the transmission of binary data from the application, thus offering a service akin to a classic network where bits are transmitted according to a SLA expressed in network metrics. In our evaluations, we observe that the task is extremely sensitive to binary errors in the text and float representations, with BERs as low as $10^{-3}$ exhibiting accuracy performances comparable to that of random models, thus advocating for error correction mechanisms. Additionally, this is a realistic assumption in a network agnostic of the data content. In these scenarios, we assume an ideal network, \emph{i.e.} error-free transmission of bits at the minimum theoretical cost, which is the BSC capacity: $C(p_e) = 1 - H_{BSC}(p_e)$ with $H_{BSC}(p_e) = -p_e \log_2(p_e) - (1-p_e)\log_2(1-p_e)$.
        
        Scenario 1 corresponds to the typical case of transmitting text over a standard network. The actual text processing is carried out at the recipient application, eventually including the conversion to sentence embeddings before classification. In the considered Bitext dataset, the customer requests are 46.89 characters long on average, resulting in 375.1 bits per message using UTF-8 for encoding. 
        
        Scenario 2 refers to a level 0 of our semantic integration taxomony: the network is unaware of the semantic nature of the representations being transmitted. Each request is encoded using the SalesForce embedding model\cite{SFRAIResearch2024}, which outputs vectors of 4096 single precision floating point numbers, \emph{i.e.} 32 bits in IEEE-754 encoding, totaling 131072 bits per message.
        
    \subsection{Non-Arbitrary Data Transmission over a Semantic Network}
    
        Scenario 3 illustrates the level 3 of the our taxonomy, where sentence embeddings are processed according to semantic preservation metrics, and the operational parameters for transmission are selected based on the performance metrics over a benchmark representative of the target application.
    
        \subsubsection{The FQI Scheme - Semantic Preserving Transmission}
        Transmitting embeddings over a network supporting binary data transmission requires the network to convert them into discrete binary representations of predetermined size, relying on the knowledge of the embedding model. To achieve this, we propose the following Fragment-Quantize-Index (FQI) operations:

            \textbf{Fragmentation}: Fragments embedding vectors $\mathbf{E} \in \mathbb{R}^N$ into $F=N/d$ fragments $\mathbf{e} \in \mathbb{R}^d$. A $N$-dimensional vector is split into $F$ $d$-dimensional vectors.
           	
           	\textbf{Quantization}: Projects each fragment $\mathbf{e} \in \mathbb{R}^d$ onto a set of reference d-dimensional vectors ${\mathbf{c_0},\mathbf{c_1},\ldots,\mathbf{c_{k-1}}}$.
            
            \textbf{Indexing}: Maps reference d-dimensional vectors $\mathbf{c_i}$ to binary representations $\mathbf{b_i} \in \mathbb{F}_2^{b}$ with $b=log_{2}(k)$. 
            
            As depicted in Fig. \ref{fig:QuantificationIndexation}, the objective of the proposed transmission scheme is to leverage the semantic similarity conveyed by the proximity of embeddings, \emph{e.g.} in the sense of L2 distance, to construct a limited set of quantized vectors that best summarize the input semantic content.
            These quantized vectors are then mapped to binary representations whose Hamming distances try to mirror the distances between quantized vectors. The idea is that minor transmission errors in the binary representations should result in small errors within the representation space thus preserving the integrity of the semantic information. 
            
            In the present example, the quantization and indexing is learnt on sentence embeddings from a generalist text dataset, the Locutusque Ultra-textbooks dataset\cite{locutusqueultratextbooks}, as encoded by the Salesforce Embedding model. The set of reference vectors for quantization is obtained using the K-means algorithm \cite{pmlr-v70-bachem17a}. The same set is used for all fragment. The description of the indexing function is out of the scope of this paper.
                
        \subsubsection{Benchmark-based Network Configuration}
        
        The FQI scheme offers significant flexibility in the size of the transmitted binary representations. The size of the fragments $d$ and the number of bits $b$ can be adjusted, allowing the total binary size of the embeddings to be tailored to fractional values. While higher values of $d$ and $b$ imply a smaller semantic distortion, they also result in increased complexity and memory requirements. 
        This adaptability enables fine-tuning of the binary representation to meet specific channel or application performance requirements or constraints.

\begin{figure}
                \centering
                \includegraphics[scale=0.455]{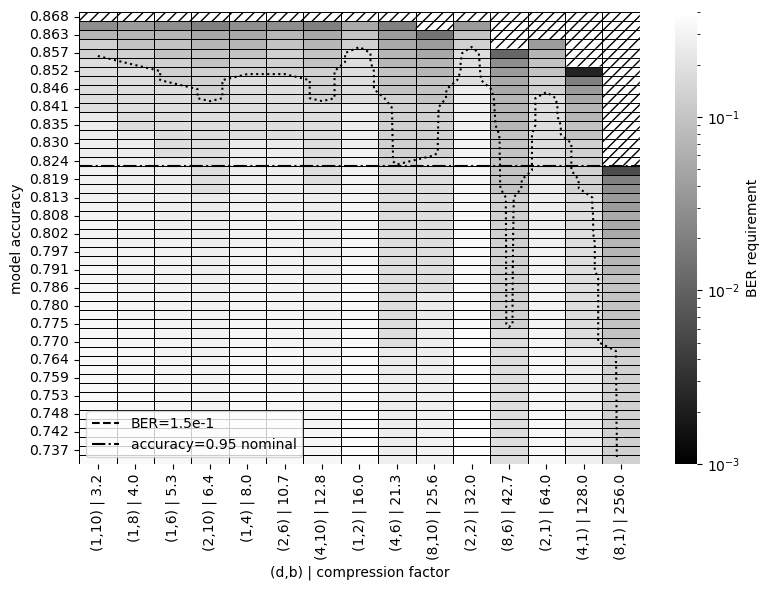}
                \caption{FQI evaluation summary: model accuracy against BER and (d,b) configurations. Nominal model accuracy is 0.8660. Acceptable accuracy loss is fixed at 95\% of nominal accuracy.}
                \label{fig:acc_vs_cf}
            \end{figure}

		To automatically adapt the network configuration, here $d$ and $b$, to the application's requirements, we propose to establish performance benchmarks on reference tasks and datasets, representative of typical applications. One such benchmark is illustrated in Fig. \ref{fig:acc_vs_cf} where the FQI configurations are evaluated on the Banking77 reference dataset and task \cite{Casanueva2020}. This evaluation includes fragment dimensions $d$ from 1 to 8, quantization levels $b$ from 1 to 10 bits, and a Bit Error Rate (BER) from $10^{-4}$ to 0.25. The FQI method's performance is reported following the Massive Text Embeddings Benchmark (MTEB) protocol \cite{muennighoff2022mteb}: ten independent evaluations are conducted for each parameter set, and for each evaluation, a logistic regressor is trained using the reference Salesforce embedding model. Importantly, the  regressor is always trained on non-quantized, error-free embeddings, which is a critical assertion due to the principle of ``separation of concern".
		
        Fig. \ref{fig:acc_vs_cf} summarizes the evaluation results for the most effective FQI configurations at a given compression factor (x-axis) across the full BER range. The model accuracy for each configuration is plotted on the y-axis, with the corresponding BER requirement represented by a color map. Incompatible configurations and requirements are shown as hashed. At a glance, it shows that the FQI can maintain benchmark-task performance comparable to the nominal model accuracy (0.8660) in a wide range of compression factors, and considering BER requirements over $10^{-3}$ that are well within the reach of a standard network. Such benchmark enables the network to be automatically configured to meet the application's demands, without having to train the network on the application's specific task, abiding again to the ``separation of concerns" principle.
        
        In our framework, the application chooses a benchmark referenced by the network similar to that of its use-case. Assuming the application's conversational agent task (Bitext) is similar to the Banking77 benchmark using the same embeddings model, the application chooses to conclude an SSLA with the network requiring a performance level (accuracy) greater than 95\% of the Banking77 nominal performance (0.8660) as a mean to meet its task requirements. Based on this agreement and observed operational conditions, the benchmark evaluation allows the network to dynamically adjust its configuration. For example, supposing a binary error rate as high as 15\%(!), delineated by dashed lines, the network can identify several $(d, b)$ configurations that still meet the SSLA requirements. In this example, the network chooses the configuration that minimizes transmission resources, \emph{i.e.} $(2,1)$ corresponding to a compression factor of 64 compared to IEEE-754 representations.
    
    \subsection{Scenarios Comparison}        
        \begin{figure}
            \centering 
            \includegraphics[scale=0.55]{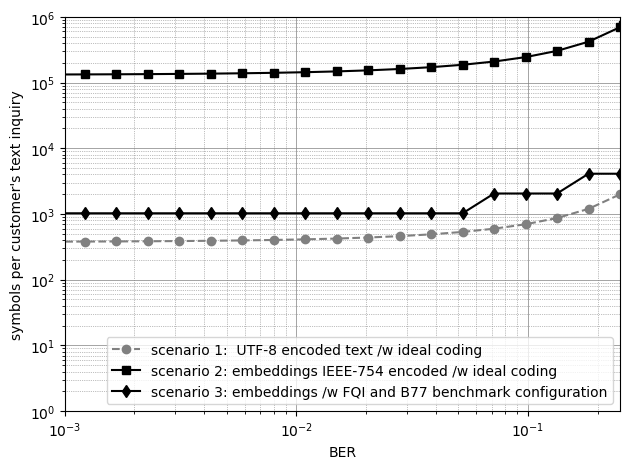} \caption{Network resources required for transmitting customer's inquiries: FQI vs arbitrary binary data transmission} \label{fig:acc_vs_ber} 
        \end{figure}

        The three presented scenarios are compared based on their utilization of network resources to accomplish the expected classification of customer queries task (Bitext task and dataset). 
        
        For scenarios 1 and 2, lacking the ability to formalize agreements based on application or benchmark performance levels, this comparison is conducted on the basis of a network ensuring error-free binary transmission, and the resources used are calculated based on the theoretical capacity of the BSC transmission channel. 
       
        For scenario 3, the resources used for the transmission of embeddings are deduced from the operating parameters $(d, b)$ derived from the benchmark used to contract the expected performance level. The FQI scheme is adapted to the encountered BER, and the transmission resources used are minimized while complying with the contracted SSLA.

        The results, expressed in logical channel symbols per customer's inquiry are presented in Fig.\ref{fig:acc_vs_ber}. Firstly, we observe a significant efficiency gain achieved through the cooperation between the network and the application in terms of transmission efficiency using the FQI scheme. The number of resources is reduced by two orders of magnitude (128 times) compared to the transmission of the same embeddings encoded in IEEE-754 binary form. 

        Additionally, we observe that the semantic integration within the network greatly mitigates the overhead of embedding transmission, while still adhering to the key principle of ``separation of concerns". In fact, we observe that the incurred overhead is only a factor of 2 to 3 compared to a UTF-8 encoded text transmission. This result is particularly noteworthy considering the assumption that transmission of the two first scenarios occurs at the theoretical capacity of the channel, and the pragmatic choice to train the FQI scheme on a general dataset rather than one specifically tailored to the targeted task.

\section{Conclusion and Perspectives}

    This work has introduced a framework for integrating semantic communication concepts into operated network, which have traditionally prioritized network-centric performance metrics over application-centric ones. The efficacy of these ideas has been demonstrated through the FQI scheme, which has yielded promising outcomes in preserving task effectiveness while optimizing the use of network resources. Future works include further studies on the FQI operations, in particular the optimization of quantization and indexing w.r.t task-specific metrics and datasets, detailed analysis of integration scenarios and the definition of information exchanged on the semantic interface. Since greater benefits are expected with data-intensive applications, they should be further investigated. Importantly, privacy issues induced by sharing information between the application and network also need to be addressed.
    
\bibliographystyle{IEEEtran}
\bibliography{references}

\end{document}